# Artificial Intelligence and the Future of Psychiatry: Insights from a Global Physician Survey

*P. Murali Doraiswamy[1,2], Charlotte Blease[3], Kaylee Bodner[1]*

*Departments of Psychiatry and Behavioral Sciences[1] and Medicine[2],
Duke University School of Medicine*

*Department of General Medicine and Primary Care[3],
Beth Israel Deaconess Medical Center, Harvard Medical School,
School of Psychology[3], University College Dublin*

*Correspondence to:*

*Professor Murali Doraiswamy, DUMC Box-3018, Durham, NC 27710, USA*

**Abstract**

**Background:** Futurists have predicted that new autonomous technologies, embedded with artificial intelligence (AI) and machine learning (ML), will lead to substantial job losses in many sectors disrupting many aspects of healthcare. Mental health appears ripe for such disruption given the global illness burden, stigma, and shortage of care providers.

**Objective:** To characterize the global psychiatrist community's opinion regarding the potential of future autonomous technology (referred to here as AI/ML) to replace key tasks carried out in mental health practice.

**Design:** Cross sectional, random stratified sample of psychiatrists registered with Sermo, a global networking platform open to verified and licensed physicians.

**Main outcome measures:** We measured opinions about the likelihood that AI/ML tools would be able to fully replace – not just assist – the average psychiatrist in performing 10 key psychiatric tasks. Among those who considered replacement likely, we measured opinions about how many years from now such a capacity might emerge. We also measured psychiatrist's perceptions about whether benefits of AI/ML would outweigh the risks.

**Results:** Survey respondents were 791 psychiatrists from 22 countries representing North America, South America, Europe and Asia-Pacific. Only 3.8% of respondents felt it was likely that future technology would make their jobs obsolete and only 17% felt that future AI/ML was likely to replace a human clinician for providing empathetic care. Documenting and updating medical records (75%) and synthesizing information (54%) were the two tasks where a majority predicted that AI/ML could fully replace human psychiatrists. Female- and US-based doctors were more uncertain that the benefits of AI would outweigh risks than male- and non-US doctors, respectively. Around one in 2 psychiatrists did however predict that their jobs would be substantially changed by AI/ML.

**Conclusions:** To our knowledge, this is the first global survey to seek the opinions of physicians on the impact of autonomous AI/ML on the future of psychiatry. Our findings provide compelling insights into how physicians think about AI/ML which in turn may help us better integrate technology and reskill doctors to enhance mental health care.

**Introduction**

Mental health disorders are estimated to affect 10-15% of the population and are among the leading causes of morbidity and mortality worldwide (1, 2). By 2030, this health burden is forecast to cost the global economy some $16 trillion (1). Suicide is the second or third leading cause of death among youth in most countries (1). Mental disorders of aging are also on the rise with the numbers of people with dementia expected to triple in coming decades. Stigma, low funding, and an acute shortage of mental health professionals (1, 2) are some of the key barriers to addressing global mental health needs. In the US, one estimate showed that 77% of counties are underserved by psychiatrists (3). In developing countries, the situation is worse - the World Health Organization estimates the rate of psychiatrists in low income countries is some 100 times lower than that in high income countries (2). India, with a population of 1.3 billion, has only about 9,000 psychiatrists (4). Addressing these large diagnostic and treatment gaps is a global public health priority (1, 2).

Against these challenges, the rapid spread of smartphones, wearable sensors, cloud based computing and intelligent technologies has led to unprecedented opportunities for patient self-monitoring, and scaling access to health care – and millions of people are already turning to such technologies (1, 5-15). The rapid pace of deep learning advances has led some experts to predict that AI is poised to disrupt healthcare and the work of doctors (6, 11, reviewed in 9). Deep learning systems can already match or outperform radiologists and pathologists in diagnostic accuracy, under controlled settings (9). Indeed, some technology futurists argue that advancements in AI/ML may one day obviate the need of physicians altogether (7, 11). Other informaticians and AI experts are less sanguine, forecasting that the role of doctors can never be fully replaced, and that the future of medicine will likely become a "team sport" between humans and machines (10, 13). Consistent with the latter view, a labor market report from Oxford University predicted that while 47% of total US employment was at risk for substitution by intelligent technology over the next two decades, the work performed by doctors would be at lower risk for automation (16). This view is also shared by authors of a more recent working paper from the Office of Economic Cooperation and Development (17) which looked at skills data from over 30 countries.

Despite lively and ongoing debate, limited attention has been paid to the views of practicing physicians on the impact of AI on medical professions. This is especially relevant in mental health care which depends on long-term, empathetic relationships between physicians and highly vulnerable patients, and in light of the flood of mental health apps available for download.

In this survey, we sought to investigate the opinions of psychiatrists about the impact of autonomous intelligent technologies, referred to as AI/ML, on the future of their jobs, as well as their potential risks and benefits in the context of mental health.

**Methods**

**Study sample**

The survey and data analyses were conducted in Spring/Summer of 2019. This was a cross-sectional global survey of psychiatrists registered with Sermo, a secure digital (online) platform designed for physician networking and anonymous survey research.

The platform is exclusive to verified and licensed physicians and has over 800,000 registered physicians, of all specialties, worldwide. The survey randomly sampled registered psychiatrists to get representation from the US, Europe and the rest of the world. As such this was an exploratory study and we aimed for a target sample size of approximately 750 psychiatrist respondents, to approximate a prior report (18). The survey collected information on nationality, demographics, perceptions of the future of psychiatry; perceptions of the workforce; and practice characteristics. The anonymous survey results were de-linked to respondent's personal identifiable information to create de-identified data. This research does not include any sensitive or identifiable data in agreement with Duke University's institutional review board requirements for exempt research.

**Survey instrument**

We used an instrument previously shown to have utility (18, 19) and modified it for this survey to include 10 questions specifically geared to key tasks that are a routine part of a psychiatrist's role. The questions were included after consultations with psychiatrists and initial piloting to ensure face validity and feasibility.

As this was a global survey, the 10 tasks were universal to psychiatrists across different countries and different types of health systems. The tasks included: *1) provide documentation (e.g. update medical records) about patients, 2) perform a mental status examination, 3) interview psychiatric patients in a range of settings to obtain medical history, 4) analyze patient information to detect homicidal thoughts, 5) analyze patient information to detect suicidal thoughts, 6) synthesize patient information to reach*

*diagnoses, 7) formulate personalized medication and/or therapy treatment plans for patients, 8) evaluate when to refer patients to outpatient versus inpatient treatment, 9) analyze patient information to predict the course of a mental health condition (prognoses), and 10) provide empathetic care to patients.* Task descriptions employed neutral language that was not biased in favor of either human physicians or technology.

To avoid potential ambiguities in how respondents interpreted the questions and response options, we focused on whether tasks were likely to be fully - rather than partially - outsourced to technology. We also aimed to allow respondents to express discriminatory opinions about the tasks they considered most (or indeed least) vulnerable to replacement by machine learning. Finally, because the term "machine learning" may be unfamiliar among some physicians and considered too narrow a description among medical AI researchers, we employed generic language such as "machines" and "technology" to refer to AI innovations.

The first set of ten items opened with a brief statement: "*Some people believe that current and future innovations in artificial intelligence will lead to significant changes in psychiatric practice and that machines will one day replace the work of psychiatrists. Others deny that new technologies will ever have the capacity to replace this work*". We then asked respondents their opinion on the likelihood that "*future technology will be able to fully replace and not merely aid human doctors in performing each task as well as or better than the average Psychiatrist.*" Employing 6-level Likert items we included the following response options: "extremely unlikely", "unlikely", "somewhat unlikely", "somewhat likely", "likely, and "extremely likely". We avoided "don't know", "neutral" or "no opinion" options on the grounds that respondents often conflate these answers (20).

Furthermore, inclusion of these choices may have precluded measurement of substantive opinions among psychiatrists: research indicates that this is a risk in self-administered questionnaires where time-pressured individuals may invest less effort in their answers (21). Participants who responded that replacement was "somewhat likely", "likely" or "extremely likely" were asked a follow-up question about how soon in their estimation technology would have the capacity to perform the task and provided with a list of five response options (0-4 years from now, 5-10 years, 11-25 years, 25-50 years, more than 50 years from now).

**Data and Statistical methods**

The de-identified data was analyzed to extract summary statistics and 95% confidence intervals. Descriptive statistics were used to examine physicians' characteristics and opinions about the likelihood of future technology replacing doctors on the ten key psychiatric tasks. We also collapsed into positive (for "somewhat likely", "likely" or "extremely likely" responses) versus negative (for "somewhat unlikely", "unlikely" or "extremely unlikely") opinions for some contrasts. Contrasts whose 95% CIs did not overlap were viewed as qualitatively different.

**Results**

**Sample characteristics**

Table 1 summarizes the demographic characteristics. The final respondent sample consisted of 791 psychiatrists representing 22 countries in North and South America,

Europe, and Asia-Pacific (Table 2). About 40% of the participating physicians were under the age of 44 years and another 34% were over the age of 55. Women comprised about 30% of the sample. About two-thirds (64%) were white with the rest describing themselves as Asian, Black, Hispanic or mixed. Participants worked in public clinics (52%), private practice (35%) and academia (13%). Most reported seeing more than 10 patients per average day.

**Opinions about AI/technological replacement of physician jobs**

About 48.7% of respondents felt that AI/ML would have no influence or only minimal influence on the future work of psychiatrists over the next 25 years (Figure 1). Only 3.8% of respondents felt it was likely that future technology would make their jobs obsolete. Another 47% predicted that their jobs would be moderately changed by AI/ML over the next 25 years (Figure 1).

**Opinions about AI/technological replacement of specific psychiatric tasks**

Results for the 10 specific psychiatric tasks are shown in Tables 3 and 4. Most respondents (83%) felt it unlikely that future technology would ever be able to provide empathic care as well as or better than the average psychiatrist. Likewise, most psychiatrists considered it unlikely that future technology could ever fully replace psychiatrists for mental status examinations (67%), evaluating acute homicidal thoughts (58%), interviewing patients in a range of settings to obtain medical history (58%), deciding whether to refer to inpatients versus outpatient treatment (55%), formulating personalized patient treatment plans (53%) or evaluating suicidal thoughts (52%).

In contrast, the majority of respondents (83%) judged it likely that future technology would be able to replace human physicians on the task of documentation (e.g. update medical records). Almost half (47%) predicted that such wherewithal would emerge in the next four years, with an additional 37% giving an estimate of 5 to 10 years from now (Figure 2). Further, a slim majority (54%) believed it likely that future technology would be able to fully replace human physicians when it comes to synthesizing information to reach diagnoses. About (32%) predicted that such wherewithal would emerge in the next four years, with an additional 41% giving an estimate of 5 to 10 years from now (Figure 3). Table 5 depicts the timeline predicted by psychiatrists for AI/ML capacity to emerge to replace them on each specific psychiatric task.

**Opinions on Potential Benefits and Risks of Future Technologies/AI**

The risk benefit judgments also varied by physician gender and practice location. Among all respondents, 40% said they were uncertain that the potential benefits of AI/ML would outweigh the possible risks/harms and another 25% said the potential benefits would not outweigh the possible risks (Figure 4). Only 36% felt that the potential benefits of future AI/ML would outweigh the possible risks in their field.

Female psychiatrists (48%) were more likely to be uncertain that the benefits of AI/ML in psychiatry would outweigh the risks than male psychiatrists (35%). Only 23% of women predicted that the benefits of AI would outweigh the possible risks compared to 41% of men (Figure 5).

Likewise, US-based (46%) were more likely to be uncertain that the benefits of AI/ML in psychiatry would outweigh the risks than those in Europe (37%) or rest of the world (32%). Only 30% of US-based psychiatrists predicted that the potential benefits of future technologies/AI would outweigh the possible risks (Figure 6).

**Predictions about how AI/ML technologies could help or harm clinical care**

Respondents were also invited to submit open ended qualitative comments to elaborate on their survey choices. These qualitative comments will be the subject of a separate article but selected comments are presented here to illustrate the level of insights.

Potential benefits of AI/ML noted (Table 6) include "eliminating human error", work under any environment", "standardization and personalization of care plans", "more efficient integration of big data", "scalability of treatment will be a big help where there is a shortage of psychiatrists", "more truthful responses from patients", "training beginner psychiatrists", "streamlining workflow to free up psychiatrist time" and "elucidating etiologies of brain diseases that are currently opaque to humans".

Possible risks/harms (Table 7) identified include "lack of empathy", "will remain a thing and not a person", "feelings of antipathy due to job displacement", "less privacy and more fatalism", "not sure how talking to AI would help stigma", "won't be able to assess mental status comprehensively", "greater burnout if administrators use AI to increase workload", and "physicians will forsake creative thinking".

**Discussion**

The World Economic Forum's 2019 report titled "*Empowering 8 Billion Minds*" highlighted that "the burden of mental illness, in terms of human suffering, is both catastrophic and growing" and that "in the 36 largest countries where treatment is not accessible to everyone, mental health conditions have resulted in over 12 billion days of lost productivity" (2).   It noted that mental health focused apps are among the fastest growing sectors in the global digital health market and called for the adoption of technologies, in an ethical, empathetic and evidence based manner, to enable better mental health for all (2).

In that context, our survey provides the first global insight into how practicing psychiatrists think about the possible benefits and risks of future intelligent technologies in mental health care as well as how AI may impact their own jobs.  Several key findings emerged.  While about 1 in 2 psychiatrists believed that AI/ML would substantially change their jobs, only 3.8% felt that it would make their jobs obsolete.  Our survey also revealed that doctors were skeptical that future technologies could perform most complex psychiatric tasks as well as or better than human doctors.  In particular, an overwhelming majority (83%) of respondents felt it unlikely that future technology would ever be able to provide empathic care as well as or better than the average psychiatrist.  The mental status examination, evaluation of dangerous behavior and formulation of a personalized treatment plan, all essential roles of a psychiatrist, were also felt to be tasks that a future AI/ML technology would be unlikely to perform as well.

We speculate there may be several explanations for the skepticism expressed by the doctors. One possibility is they are cautious of the hype around AI (22), especially given AI's many boom and bust cycles over the past five decades, and placing high value on human interaction and personalized professional analysis. While a growing number of research studies have documented the utility of AI tools for mental health diagnosis and care (2,9,14), we are still far away from an AI that accurately recognizes and understands the full range of human emotions and mental illnesses. Further, unlike pattern-based fields like radiology or pathology where AI can sometimes outperform doctors, psychiatry requires greater integration of cultural and psychosocial factors with medical comorbidities. The contrasting explanation for the survey findings could be that doctors may be overvaluing their skills (9) and/or underestimating the rapid pace of progress in intelligent technologies. If the latter is true, it also raises questions about the preparedness of the profession to navigate technological change in the delivery of patient care (2, 9, 11, 12, 14). The range of opinions obtained from respondents suggest that multiple conflicting beliefs and factors may be at play.

The validity of our survey findings is supported by recent research (16-18, 23). A highly cited labor market report by Oxford professors Carl Frey and Michael Osbourne looked at the risk of displacement due to automation for 702 occupations (16). They rated the jobs on skills needed such as perception, creative intelligence, social intelligence, empathy, and manual dexterity. While 47% of the total US labor market was felt to be at risk for job losses, the work of mental health social workers and physicians were felt to be at lower risk (16). A projection of the Office of Economic Cooperation and Development, using data from over 30 countries, also found that physician jobs would be at low risk for automation (17). A 2018 UK survey found that family physicians believed it was unlikely that machines could replace them in delivering empathic care (18). A survey which examined Korean doctors' knowledge of AI found that only 35% agreed that AI could replace doctors in their jobs (23).

Another key insight that emerged from our survey was that a relatively high number of physicians (40%) were uncertain that the possible benefits of future AI in mental health would outweigh the possible risks/harms. In particular more female psychiatrists (versus males) and more US psychiatrists (versus those elsewhere) were uncertain that the benefits of AI would outweigh the potential risks. We do not know the reasons for this but there are some speculative possibilities including the fact that doctors may not feel confident in their knowledge of AI and may find it difficult to separate marketing hype from ground truth.

The gender differences in AI risk perception noted in our survey are novel but may be commensurate with a large body of findings that women are more risk averse than men (24). Thus, female psychiatrists may be more cautious and circumspect in weighing up the benefits versus harms of AI/ML, especially where ambiguities persist with respect to ethics, biases, inequities, data privacy and risks of poorly validated "black box" algorithms (2, 10, 25). Unlike European countries which operate with universal health coverage and strict regulations about consumer and citizen data privacy, the US operates on multiple insurance-systems and has substantially weaker data privacy rules. This may be one reason that US health professionals perceive patients to be at graver risk from the sharing of health information gathered by electronic devices and continuous online monitoring in being sold to third parties and in determining private-insurance policies. The EU General Data Protection Regulation (GDPR) gives control to individuals over their personal data and requires businesses to use the highest-possible privacy settings by default.  While parallel legislation may come into effect in California in January 2020, no such laws have so far been enacted at a federal level in the USA – nor is there much appetite to do so.  Lack of legislative movement on citizen protections in respect of health data gathering may have further aroused caution about the benefits of machine learning among vulnerable patients, in the US contexts.

Practitioners based in several middle and lower income countries expressed more optimism than those in the USA about the beneficial impact of AI on professional practice – possibly due to lower access to care in low and middle income settings (1). As a recent report from the World Health Organization noted the shortage of psychiatrists in low income countries is some 100-fold greater than it is in the wealthy nations (1). The potential for AI/ML innovations to optimize efficiency via the provision of cheaper, scalable access to mental health care, may have influenced the views of psychiatrists working in under-resourced health systems, where stigmatization of mental health conditions may also be a key, ongoing consideration.

Last but not least, our survey provides powerful insights from a global sample of psychiatrists – the end users – into how AI/ML could be optimally deployed to work with physicians, rather than replace them, to enhance mental health care.  A crucial reason that new technologies fail to be adopted in healthcare is that they often do not take the end user into account.  Respondents identified ways that their effectiveness and face time with patients could be improved by AI/ML as well as several potential risks/harms from future AI – all of which may be informative to technology developers, regulators, payers and consumers. While doctors were skeptical about the prospects of AI/ML replacing them, about one in two psychiatrists felt that future technologies would significantly transform their jobs.  Psychiatrists also predicted that AI/ML could aid in several ways such as reducing administrative burden, 24/7 monitoring, individualized drug targets to reduce side effects, integration of new streams of data from wearables and genetics, reducing human errors, scaling care to areas with psychiatrist shortage and elucidating brain etiologies that are currently opaque.  Concerns and risks identified include loss of privacy, lack of transparency, unknown effects on stigma, incorrect diagnosis or treatment, dehumanization, lack of empathy, greater physician burnout due to higher throughput and loss of control.  These findings, which add to those noted by experts (2, 9, 10, 14, 25), should be a priority for further research and ethical oversight.

**Strengths and Limitations**

This is the first global survey to investigate the opinions of psychiatrists about the impact of AI on their profession. The survey benefited from a relatively large sample of psychiatrists drawn from diverse practice settings across 22 countries.  In addition, employing Sermo, a global platform of verified and licensed physicians, allowed us to recruit frontline practicing psychiatrists.  There are some limitations with our survey. First, was the relative absence of respondents from developing nations (such as in Africa), sampling bias (e.g. people registered on a platform), response biases (e.g. degree to which they are interested in topic may have influenced their participation) as well as confounding effects of variables not measured. In the absence of a population margin of error estimate, we computed a 95% confidence interval. We focused on broad psychiatric functions rather than granular subtasks to reduce anthropocentric bias. Lastly, as with many surveys, we cannot determine causality or predictive validity – the impact of machine learning on psychiatry may not be known for decades. Hence, our findings must be interpreted within this context as preliminary. A population survey to replicate and expand on our findings is warranted. Despite such limitations, this global survey provides foundational insights into how psychiatrists think about future technologies in mental health care.

**Conclusions**

Future surveys might usefully examine the views of patients, and those suffering from mental illness on the impact of AI on psychiatry and mental health services; and on forecasts among informaticians and AI experts working at the nexus of mental health research and practice. Combined, such insights would collectively help to better develop and validate, and better prepare mental health professionals and patients to implement machine learning technologies to address this most vital challenge to global health and wellbeing.

**Acknowledgments**

The authors would like to thank Sermo, especially Peter Kirk and Joanna Molke, for their collaboration. We would also like to express our gratitude to the physicians who participated in this survey and shared their valuable insights. There was no external funding support for this study. Dr Doraiswamy has received research grants from and/or served as an advisor or board member to government agencies, technology and healthcare businesses and advocacy groups for other projects in this field. Dr Blease was supported by an Irish Research Council-Marie Skłodowska-Curie Global Fellowship.

**Table 1. Demographics of the Psychiatrists (N=791)**

| Gender | % |
| --- | --- |
| Male | 69.5% |
| Female | 29.2% |
| Prefer not to say | 1.1% |
| **Age** | |
| 25-34 | 9.7% |
| 35-44 | 29.3% |
| 45-54 | 26.7% |
| 55-64 | 24.7% |
| 65 and above | 9.6% |
| **Race/ethnicity** | |
| Asian | 17.6% |
| Black/African/Caribbean | 2.0% |
| Mixed/Multiple Ethnic Groups | 3.7% |
| White | 64.3% |
| Other Ethnic Group Not Listed | 3.2% |
| Prefer not to say | 9.3% |

**Table 2. Country where the Respondent Psychiatrist Practices**

| Country | N | % |
|---|---|---|
| UNITED STATES | 276 | 34.9% |
| FRANCE | 77 | 9.7% |
| ITALY | 74 | 9.4% |
| GERMANY | 59 | 7.5% |
| SPAIN | 57 | 7.2% |
| UNITED KINGDOM | 50 | 6.3% |
| RUSSIAN FEDERATION | 30 | 3.8% |
| AUSTRALIA | 25 | 3.2% |
| JAPAN | 22 | 2.8% |
| MEXICO | 20 | 2.5% |
| CANADA | 18 | 2.3% |
| GREECE | 15 | 1.9% |
| CHINA | 14 | 1.8% |
| BRAZIL | 12 | 1.5% |
| POLAND | 11 | 1.4% |
| TURKEY | 11 | 1.4% |
| NETHERLANDS | 8 | 1.0% |
| BELGIUM | 4 | 0.5% |
| SWITZERLAND | 3 | 0.4% |
| NORWAY | 2 | 0.3% |
| PORTUGAL | 2 | 0.3% |
| INDIA | 1 | 0.1% |

**Table 3. Responses to the question "in your opinion what is the likelihood that future technology will be able to replace human doctors to perform these tasks as well as or better than the average psychiatrist?"**

| Task | Opinions Frequency-percentage (95% CI) | | | | | |
|---|---|---|---|---|---|---|
| | Extremely unlikely | Unlikely | Somewhat unlikely | Somewhat likely | Likely | Extremely likely |
| Provide documentation (e.g., update medical records) about patients | 6 (4.3 to 7.7) | 8 (6.1 to 9.9) | 11 (8.8 to 13.2) | 24 (21.0 to 27.0) | 24 (21.0 to 27.0) | 28 (24.9 to 31.1) |
| Provide empathetic care to patients | 53 (49.5 to 56.5) | 19 (16.3 to 21.7) | 11 (8.8 to 13.2) | 9 (7.0 to 11.0) | 5 (3.5 to 6.5) | 3 (1.8 to 4.2) |
| Formulate personalized medication and/or therapy treatment plans for patients | 16 (13.5 to 18.6) | 20 (17.2 to 22.8) | 17 (14.4 to 19.6) | 25 (22.0 to 28.0) | 14 (11.6 to 16.4) | 8 (6.1 to 9.9) |
| Evaluate when to refer patients to outpatient versus inpatient treatment | 17 (14.4 to 19.6) | 19 (16.3 to 21.7) | 19 (16.3 to 21.7) | 26 (22.9 to 29.1) | 12 (9.7 to 14.3) | 7 (5.2 to 8.8) |
| Analyze patient information to establish prognoses | 15 (12.5 to 17.5) | 15 (12.5 to 17.5) | 19 (16.3 to 21.7) | 27 (23.9 to 30.1) | 16 (13.5 to 18.6) | 8 (6.1 to 9.9) |
| Analyze patient information to detect acute homicidal thoughts | 20 (17.2 to 22.8) | 21 (18.2 to 23.8) | 17 (14.4 to 19.6) | 23 (20.1 to 25.9) | 12 (9.7 to 14.3) | 6 (4.3 to 7.7) |
| Analyze patient information to detect suicidal thoughts | 18 (15.3 to 20.7) | 17 (14.4 to 19.6) | 16 (13.5 to 18.6) | 25 (21.9 to 28.0) | 14 (11.6 to 16.4) | 9 (7.0 to 11.0) |
| Synthesize patient information to reach diagnoses | 15 (12.5 to 17.5) | 15 (12.5 to 17.5) | 16 (13.5 to 18.6) | 29 (25.8 to 32.2) | 17 (14.4 to 19.6) | 9 (7.0 to 11.0) |
| Perform a mental status examination | 27 (23.9 to 30.1) | 24 (21.0 to 27.0) | 16 (13.5 to 18.6) | 17 (14.4 to 19.6) | 10 (7.9 to 12.1) | 6 (4.3 to 7.7) |
| Interview psychiatric patients in a range of settings to obtain medical history | 24 (21.0 to 27.0) | 21 (18.2 to 23.8) | 13 (10.7 to 15.3) | 20 (17.2 to 22.8) | 14 (11.6 to 16.4) | 8 (6.1 to 9.9) |

**Table 4. Responses to the question "in your opinion what is the likelihood that future technology will be able to replace human doctors to perform these tasks as well as or better than the average psychiatrist?"***

| Task | Opinions Frequency-percentage (95% CI) | |
|---|---|---|
| | **Unlikely** | **Likely** |
| Provide documentation (e.g., update medical records) about patients | 25 (22.0 to 28.0) | 75 (72.0 to 78.0) |
| Provide empathetic care to patients | 83 (80.4 to 85.6) | 17 (14.4 to 19.6) |
| Formulate personalized medication and/or therapy treatment plans for patients | 53 (49.5 to 56.5) | 47 (43.5 to 50.5) |
| Evaluate when to refer patients to outpatient versus inpatient treatment | 55 (51.5 to 58.5) | 45 (41.5 to 48.5) |
| Analyze patient information to establish prognoses | 49 (45.5 to 52.5) | 51 (47.5 to 54.5) |
| Analyze patient information to detect acute homicidal thoughts | 58 (54.6 to 61.4) | 42 (38.6 to 45.4) |
| Analyze patient information to detect suicidal thoughts | 52 (48.5 to 55.5) | 48 (44.5 to 51.5) |
| Synthesize patient information to reach diagnoses | 46 (42.5 to 49.5) | 54 (50.5 to 57.5) |
| Perform a mental status examination | 67 (63.7 to 70.3) | 33 (29.7 to 36.3) |
| Interview psychiatric patients in a range of settings to obtain medical history | 58 (54.6 to 61.4) | 42 (38.6 to 45.4) |

*The 6 Likert categories have been combined into two categories (likely and unlikely).

**Table 5. Predicted timespan when AI/ML technology will have the capacity to replace human physicians for specific clinical tasks***

| | Predicted timespan<br>*Frequency-percentage*<br>*(95% CI)* | | | | |
|---|---|---|---|---|---|
| **Task** | **0-4 years** | **5-10 years** | **11-25 years** | **26-50 years** | **> 50 years** |
| Provide documentation (e.g., update medical records) about patients | 47<br>(42.2 to 51.8) | 37<br>(32.3 to 41.7) | 11<br>(8.0 to 14.0) | 3<br>(1.4 to 4.7) | 2<br>(0.8 to 3.4) |
| Provide empathetic care to patients | 28<br>(16.9 to 39.1) | 36<br>(24.2 to 47.9) | 20<br>(10.1 to 29.9) | 8<br>(1.3 to 14.7) | 8<br>(1.3 to 14.7) |
| Formulate personalized medication and/or therapy treatment plans for patients | 35<br>(27.9 to 42.1) | 39<br>(31.8 to 46.3) | 18<br>(12.3 to 23.7) | 6<br>(2.5 to 9.5) | 2<br>(-0.1 to 4.1) |
| Evaluate when to refer patients to outpatient versus inpatient treatment | 33<br>(25.5 to 40.5) | 44<br>(36.1 to 51.9) | 14<br>(8.5 to 19.6) | 8<br>(3.66 to 12.3) | 1<br>(-0.6 to 2.6) |
| Analyze patient information to establish prognoses | 28<br>(21.6 to 34.4) | 49<br>(41.9 to 56.1) | 15<br>(9.9 to 20.1) | 5<br>(1.9 to 8.1) | 3<br>(0.6 to 5.4) |
| Analyze patient information to detect acute homicidal thoughts | 37<br>(29.1 to 44.9) | 40<br>(31.9 to 48.1) | 12<br>(6.7 to 17.3) | 8<br>(3.5 to 12.5) | 3<br>(0.2 to 5.8) |
| Analyze patient information to detect suicidal thoughts | 34<br>(27.1 to 40.9) | 43<br>(35.8 to 50.2) | 13<br>(8.1 to 17.9) | 8<br>(4.1 to 11.9) | 2<br>(0 to 4.0) |
| Synthesize patient information to reach diagnoses | 32<br>(25.6 to 38.4) | 41<br>(34.3 to 47.7) | 19<br>(13.6 to 24.4) | 5<br>(2.0 to 8.0) | 3<br>(0.7 to 5.3) |
| Perform a mental status examination | 31<br>(23.0 to 39.0) | 41<br>(32.5 to 49.6) | 18<br>(11.3 to 24.7) | 9<br>(4.0 to 14.0) | 2<br>(-0.4 to 4.4) |
| Interview psychiatric patients in a range of settings to obtain medical history | 27<br>(20.4 to 33.6) | 45<br>(37.6 to 52.4) | 20<br>(14.1 to 25.9) | 6<br>(2.5 to 9.5) | 2<br>(-0.1 to 4.1) |

*question only asked of respondents who felt future AI/ML would likely replace them on that task

**Table 6. Physicians' comments about the potential benefits of AI/ML in Psychiatry***

| |
|---|
| "It can work under any environment, no fatigue and not moody." "AI can be more easily controlled than us" |
| "Eliminate the human error; Less medical errors, more standardized protocols, better outcomes." |
| "Better quality of life for psychiatrist if work is streamlined. This will lead to either increase in productivity or increase in free time." |
| "The patient can answer more truthfully to artificial intelligence and to accept the encouragement-support more objectively." |
| "Making standardized plans and doing standard assessments." |
| "Less bias due to race or gender, can use big data more efficiently than humans." |
| "It might help patients at risk seek care sooner." |
| "It will be of great benefit in areas where there is shortage of psychiatrists. Scalability of treatment." |
| "Providing practical guidance for beginners psychiatrists." |
| "AI will also help us elucidate etiologies of brain diseases that are currently opaque to us." |
| "The major question appears to be whether society will tolerate use of AI as medical practitioner, rather than whether it is capable of doing..." |

*selected to illustrate the range of opinions

**Table 7. Physicians' comments about the potential harms of AI/ML in Psychiatry***

| |
|---|
| "Lack of empathy/no humanity would jeopardize therapeutic process." "Empath may be a specific job title down the road..someone who works with an AI." |
| "There may be feelings of antipathy toward AI/tech due to job displacement." |
| "AI won't be able to assess patient's mental status comprehensively, and therefore will be unable to determine the exact problem." |
| "There is a possibility that the process for the machine to reach a diagnosis could turn into a black box process, ..beyond human understanding." |
| "There is a stigma associated with mental health treatment already and am not sure how talking to an AI would help the treatment process at all." |
| "Until the machine is aware of simulating consciousness, even perfectly it will not eliminate the pain of loneliness. AI will remain a thing, not a person" |
| "A potential harm is patients having less privacy and more fatalism in general." |
| "If the time saved is used (by administrators) to increase psychiatrists' patient loads, it might lead to greater burnout." |
| "Risks of worsening dehumanization and worsening some delirious symptoms" |
| "Physicians will forsake creative clinical thinking." |

*selected to illustrate the range of opinions

**Figure 1. Predicted impact of future AI/ML on the work of psychiatrists over the next 25 years***

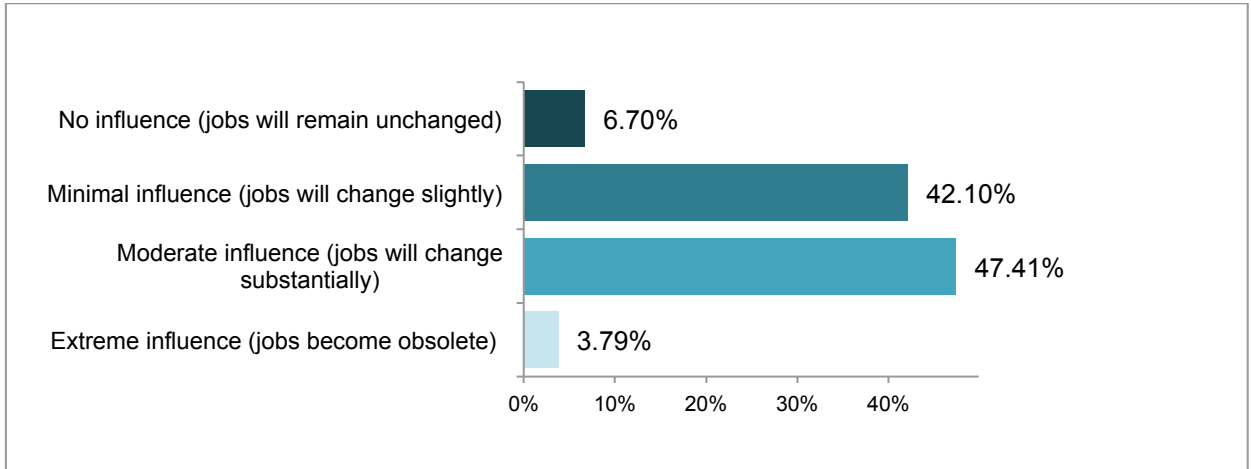

*percentages may not add up to 100% due to rounding

**Figure 2. Physicians' opinions on when technology will have the capacity to replace the average psychiatrist in providing documentation about patients**[*]

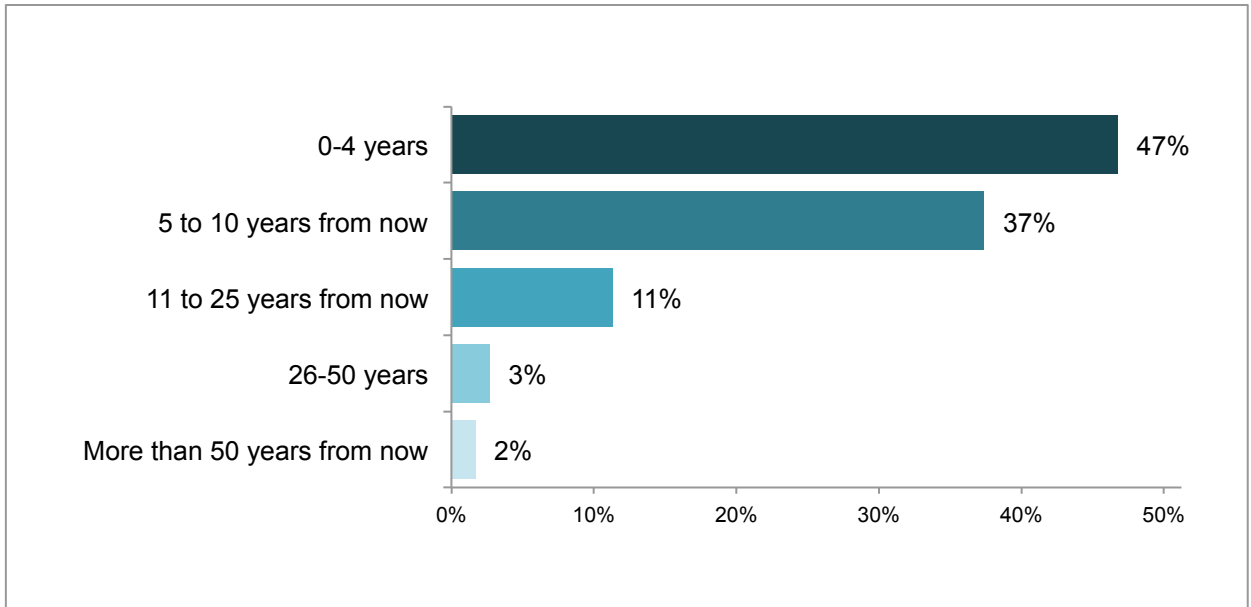

*Only includes psychiatrists who selected "Extremely likely" or "Likely"

**Figure 3. Physicians' opinions on when technology will have the capacity to replace the average psychiatrist in synthesizing patient information to reach diagnosis**[*]

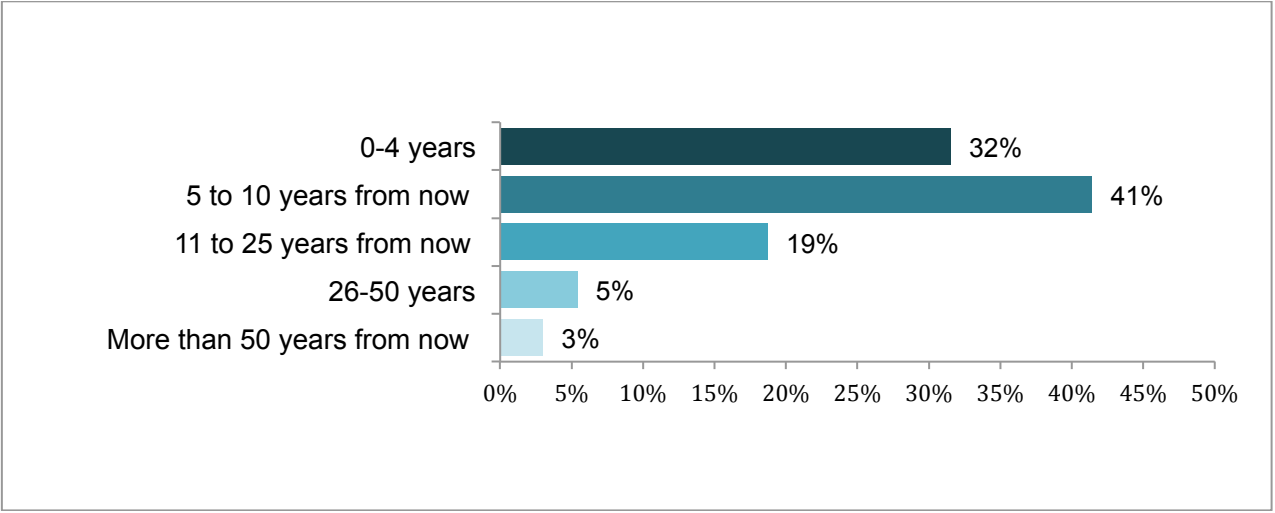

[*]only includes psychiatrists who selected "Extremely likely" or "Likely"

**Figure 4. Psychiatrists' opinions on whether potential benefits of AI outweigh possible risks/harms**[*]

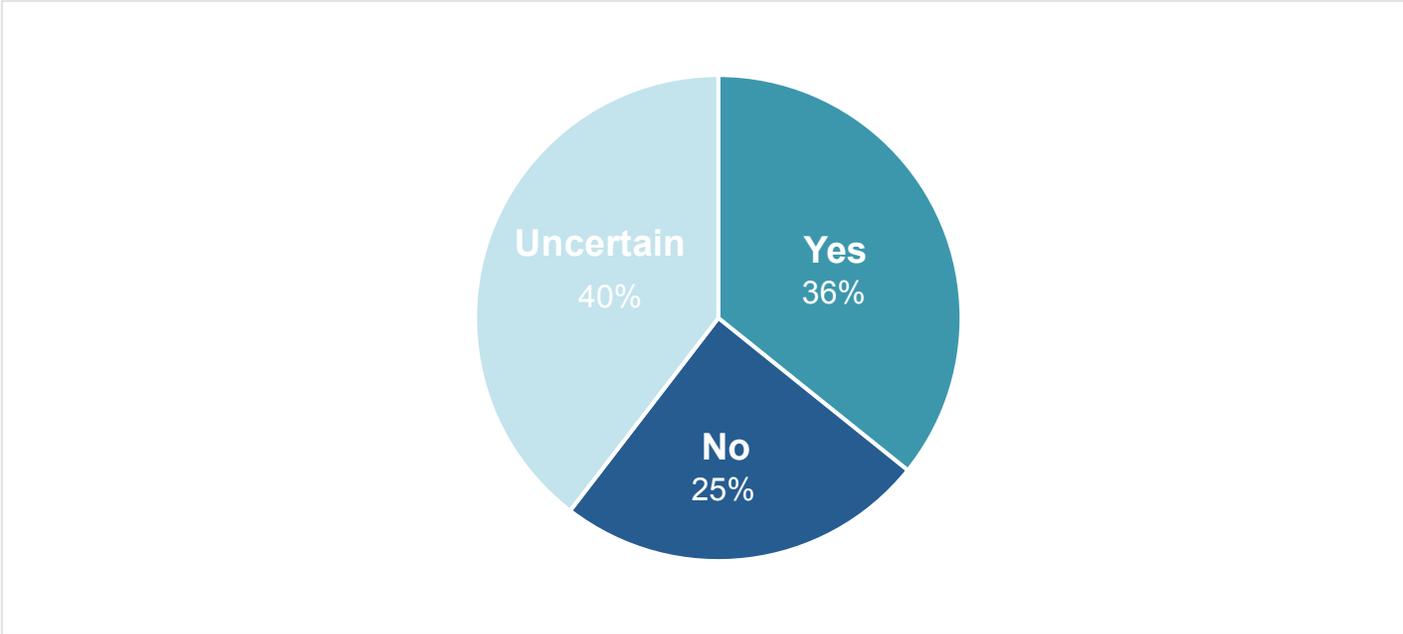

[*]percentages may not add up to 100% due to rounding.

**Figure 5. Psychiatrists' opinions on whether potential benefits of AI outweigh possible risks/harms**[*]

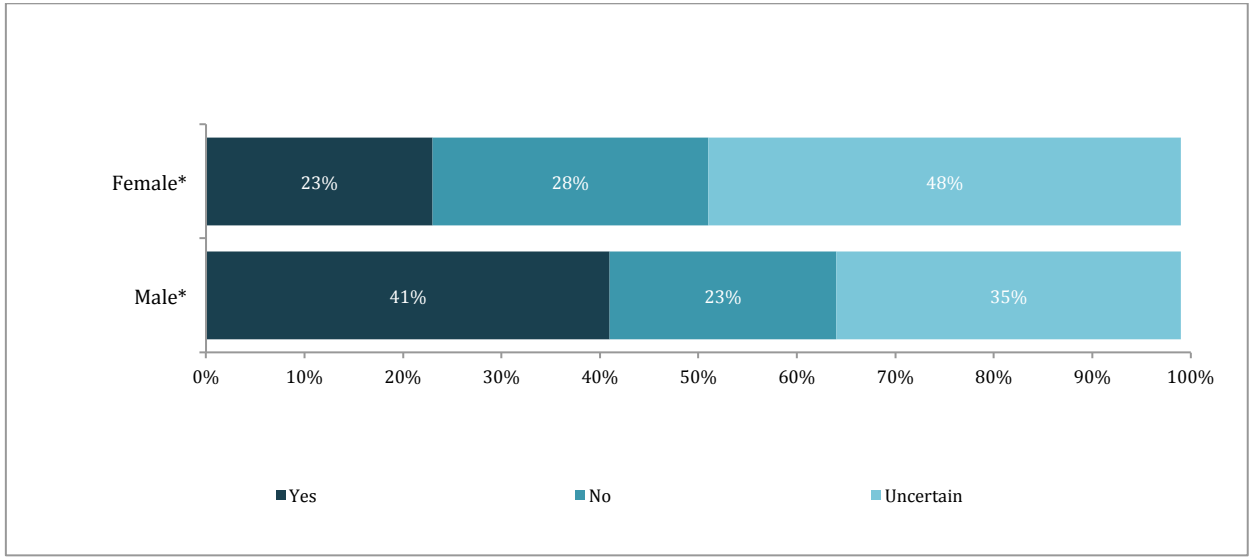

[*]data shown by gender, percentages may not add up to 100% due to rounding; some subjects did not mark gender.

**Figure 6. Psychiatrists' opinions on whether potential benefits of AI outweigh possible risks/harms**[*]

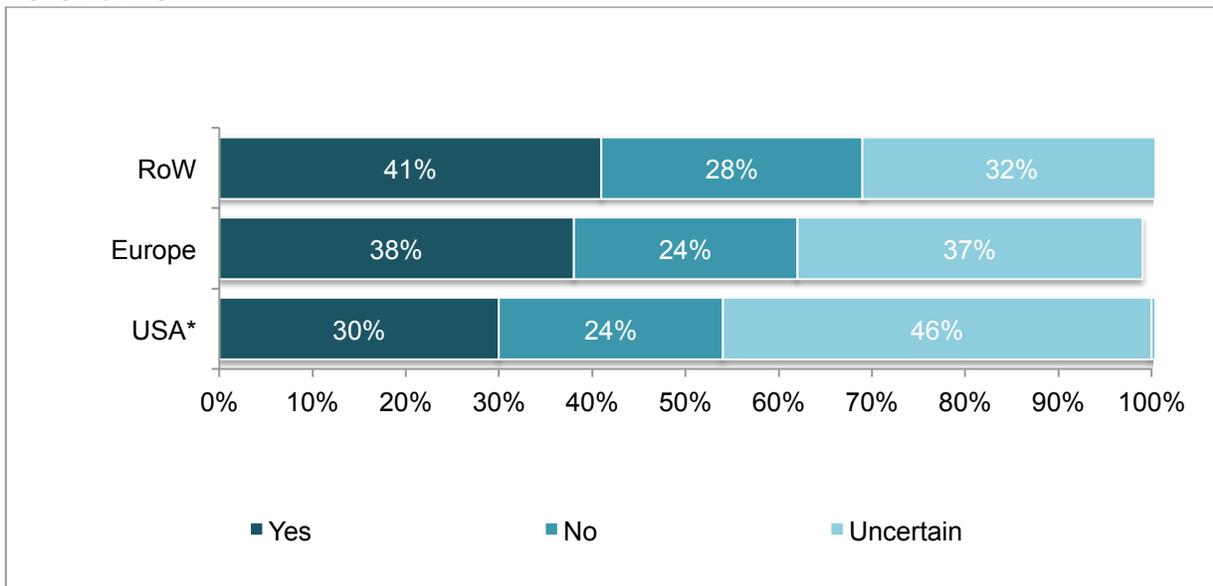

[*]data shown by region of practice; percentages may not add up to 100% due to rounding; RoW= rest of world